\documentclass[12pt]{article}
\usepackage{amsmath,amssymb,url,epsfig,color,cite}
\usepackage{hyperref}	

\newcommand{\be}{\begin{equation}}
\newcommand{\ee}{\end{equation}}
\newcommand{\beq}{\begin{equation}}
\newcommand{\eeq}{\end{equation}}

\begin{document}
\begin{center}
\hfill
CERN-PH-TH/2012-250 \\
\hfill 
ICTP-SAIFR/2012-005

\bigskip\bigskip\bigskip

{\Large\bf
Multi Higgs and Vector boson
production beyond the Standard Model}\\

\medskip
\bigskip\color{black}\vspace{0.6cm}
{\bf
\large A. Belyaev$^{1,2}$,  A.~C.~A.~Oliveira$^{3,4}$, R.~Rosenfeld$^{3,4}$
and  M.~C.~Thomas$^{1}$}
\\[7mm]
{\it $^1$ School of Physics \& Astronomy, University of Southampton, UK}\\
{\it $^2$ Particle Physics Department, Rutherford Appleton Laboratory, 
Chilton, Didcot, Oxon OX11 0QX, UK} \\
{\it $^3$ ICTP South American Institute for Fundamental Research \\
Instituto de F\'{\i}sica Te\'orica - 
Universidade Estadual Paulista\\
Rua Dr. Bento T. Ferraz 271, 01140-070,
S\~ao Paulo, Brazil } \\
{\it $^4$ Theory Division, Physics Department, CERN, CH-1211, Geneva 23, Switzerland}

\bigskip\bigskip\bigskip\bigskip

{
\centerline{\large\bf Abstract}
\begin{quote}

\medskip

If the electroweak symmetry breaking is originated from a strongly
coupled sector, as for instance in composite Higgs models, the Higgs boson
couplings can deviate from their Standard Model values. In such cases, 
at sufficiently high energies there could occur an onset of multiple Higgs boson and 
longitudinally polarised electroweak gauge boson ($V_L$) production.
We study the sensitivity to anomalous Higgs couplings in inelastic processes with 3 and 4 particles 
(either Higgs bosons or $V_L$'s) in the final state. We show that, due to the more severe cancellations
in the corresponding amplitudes as compared to the usual $2 \rightarrow 2$ processes, large enhancements
with respect to the Standard Model can arise even for small modifications of the Higgs couplings. 
In particular, we find that triple Higgs production provides the best multiparticle channel to look for these
deviations. We briefly explore the consequences of  multiparticle production at the LHC.
\end{quote}}

\end{center}

\section{Introduction}
The search for the mechanism responsible for electroweak symmetry breaking (EWSB) has been 
a long-lasting endeavour  that may finally be resolved with the
new data from the LHC. However, even with the recent evidence for a light scalar state, the jury is still out on 
whether EWSB is caused by new strong interactions or not.
In order to have a definitive answer further experimental probes are necessary.

In particular, a hallmark of strong interactions in the EWSB sector is multiple particle production \cite{CG} 
\footnote{One should note, however, that multiple particle production with large cross section could also be obtained
in weakly coupled theories at tree level simply due to the large number of diagrams, but this is expected only for
very large multiplicities, of order ${\cal O} (1/\alpha_{EW})$ \cite{Rubakov}.}. 
In a strongly coupled EWSB sector one would expect copious production of
longitudinal gauge bosons granted enough energy is available to produce them.
This would be similar to the production of events with a large pion multiplicity in QCD at high energies.
In fact, multi-W production was studied in a simplified scaled-up version of QCD almost 20 years ago \cite{MPR}. 
In this Letter we study the inelastic production of longitudinally polarised W and Z bosons (denoted collectively by $V_L$) and Higgs bosons
in the context of an effective Lagrangian. We estimate the energy scale at which these processes become relevant, which signals the onset
of new physics, as recently discussed in \cite{AAD}. In particular, we will be interested in the sensitivity to non-SM Higgs 
couplings in the growth of the cross section for these processes. Our results have as a particular case
the study of unitarity violation in multi-$V_L$ production in the Higgsless model \cite{DH}.
We show that models with partial unitarization, such as the composite Higgs model, can lead to a large
enhancement of multiparticle cross section due to the absence of
cancellation mechanisms in the corresponding scattering
amplitudes. This effect becomes more acute as the final state multiplicity increases, provided that enough energy is available.

\section{Multiparticle cross sections and unitarity}

The perturbative unitarity bound in the inelastic $2 \rightarrow n$ process assuming $s$-wave
dominance for a given center-of-mass energy $\sqrt{s}$ is \cite{DH,MNW}:
\begin{equation}
\sigma (2 \rightarrow n) < \frac{4 \pi}{s}
\label{unitarity}
\end{equation}

This unitarity bound sets stringent constraints on the scattering amplitudes. 
Since the relativistic $n-$body phase space is proportional to $s^{n-2}$ the unitarity bound requires 
that the amplitude grows with energy no faster than
\begin{equation}
A(2\rightarrow n) \sim s^{1-n/2}
\end{equation}
Suppose, for instance, that the
EWSB sector is described by a simple nonlinear sigma model (NL$\sigma$M), neglecting transverse gauge bosons for the moment, 
and assuming possible resonance states to be very heavy:
\begin{equation}
{\cal L}_{NL\sigma M} = \frac{v^2}{4} \mbox{Tr} \left[ \partial_\mu U \partial^\mu U^\dagger \right]
\end{equation}
where $v=246$ GeV is the usual scale of electroweak symmetry breaking and
\begin{equation}
U = e^{\frac{i \vec{\tau} \cdot \vec{\pi}}{v}}.
\end{equation}
The isospin triplet ``pion" fields $\pi^i$ ($i=1,2,3)$ will be identified with the longitudinally polarised gauge bosons through the equivalence theorem \cite{CG}. 
By power-counting, the scattering amplitude in this model grows with energy as
\begin{equation}
A_{NL\sigma M} (2 \rightarrow n) \sim \frac{s}{v^n}
\end{equation}
and hence naively
\begin{equation}
\sigma (2 \rightarrow n) \sim \frac{1}{s} \left(\frac{s}{v^n}\right)^2 s^{n-2}.
\end{equation}
Therefore, the growth of the cross section towards the unitarity bound in this model is faster for larger
number of particles due to the kinematical factors in the phase space, assuming of course that enough energy is available.

Conversely, there must be stronger cancellations in the scattering amplitudes due to new physics as the number 
of final state particles is increased. For instance, unitarity requires that $A(2 \rightarrow 2) \sim \mbox{constant}$ and  $A(2 \rightarrow 4) \sim 1/s$, 
whereas they both grow as $\sim s$ in the NL$\sigma$M.
Therefore, in the absence of a perfect cancellation, it is plausible that the growth of the cross section may have a
large impact in multi-$V_L$ production. It is the purpose of this work to examine this impact.


Given the fully relativistic $n-$body phase space given by \cite{BK}:
\begin{equation}
R_n(s) = \int \prod\limits_{i=1}^{n} \frac{d^3 p_i}{(2 \pi)^3 (2 E_i)^3} (2 \pi)^4 \delta^4(\sqrt{s} - \sum\limits_{i=1}^{n} p_i) =
\frac{ (2 \pi)^{4-3n} (\pi/2)^{n-1}}{(n-1)! (n-2)!} s^{n-2}
\end{equation}
one can easily estimate the energy scale $\Lambda_n$ at which perturbative unitarity is violated in $2 \rightarrow n$ processes 
in the NL$\sigma$M:
\begin{equation}
\Lambda_n = \left[ \frac{2 (n-1)! (n-2)!}{(2 \pi)^{3-3n} (\pi/2)^{n-1}} \right]^{\frac{1}{2n}}  v.
\end{equation}
For example, unitarity is violated in $2 \rightarrow 4$ processes at an energy which is almost 2.5 higher
than that for  the usual $2 \rightarrow 2$ processes. 
This estimate is in reasonable agreement with \cite{DH}.
One should notice that we are not including in this rough estimate the growth due to the combinatorial factors and a proper
phase space integration.
These will be included below in a fully automated calculation.

\section{Anomalous Higgs couplings and partial unitarization}

In order to recover unitarity there must an UV completion of the model describing the interaction
of the lightest degrees of freedom. The simplest possibility is the addition of a scalar field, which is identified
with the Higgs scalar. 
However, it is possible that the Higgs scalar is a composite particle with couplings that may differ from the SM ones.
In this case, the theory is not UV-complete and  unitarity is only partially restored.
For such a theory one can use an effective Lagrangian (SMEFF) to parameterise its 
couplings to longitudinally polarised gauge bosons and self-couplings (couplings to fermions are not relevant to the results presented here)  \cite{SILH}:
\begin{eqnarray}
{\cal L}_{\mbox{{\small eff}}}  &=& 
\frac{v^2}{4} \left(1 + 2 a \frac{h}{v} + b \frac{h^2}{v^2} + b_3 \frac{h^3}{v^3} + \cdots \right) \mbox{Tr} \left[ \partial_\mu U \partial^\mu U^\dagger \right]
\nonumber \\
&+& 
\frac{1}{2} (\partial_\mu h)^2 - \frac{1}{2} m_h^2 h^2  - d_3 \lambda v h^3 - d_4 \frac{\lambda}{4} h^4 + \cdots
\label{SILH}
\end{eqnarray}
This parameterisation is common to a large class of models, such as composite Higgs models, and has been used 
to study anomalous Higgs couplings in $V_L V_L \rightarrow V_L V_L, hh$ processes at the LHC \cite{SILH,grojeanetal,GM}. 
Unitarity is recovered for the SM values $a=b = d_3 = d_4 = 1$ and $b_3 =0$. For different values of these parameters the usual cancellation provided by the scalar field
is incomplete. As an example, the Minimal Composite Higgs Model (MCHM4) predicts that the couplings of the ``pions" with the Higgs boson follows from
an expansion around the vacuum $h(x)=0$ of the effective Lagrangian \cite{CMPR}
\begin{equation}
\frac{f^2}{4} \sin^2\left( \theta + \frac{h(x)}{f} \right) \mbox{Tr} \left[ \partial_\mu U \partial^\mu U^\dagger \right] 
\end{equation}
with the identification $v = f \sin \theta$, which comes from the mass term for the gauge fields. 
This Lagrangian has a discrete symmetry under the parity transformation $h \rightarrow -h$ and $\pi \rightarrow -\pi$, 
although this is not obvious in this representation\cite{CMPR}. 
Therefore, in the MCHM4
\begin{equation}
a = \sqrt{1 - \xi}; \;\; b = 1 - 2 \xi; \;\; b_3 = -\frac{4}{3} \xi \sqrt{1-\xi};  \;\; \cdots
\label{coeffMCHM4}
\end{equation}

In order to study the $2 \rightarrow 4$ scattering, one must expand each field $U$ to order ${\cal O}(\pi^6)$:
\begin{eqnarray}
&&\frac{v^2}{4}  \mbox{Tr} \left[ \partial_\mu U \partial^\mu U^\dagger \right]  = \frac{1}{2} \left(\partial^\mu\vec{\pi} \cdot \partial_{\mu} \vec{\pi} \right)   +
\left[ 1 - \frac{2}{15 v^2} \vec{\pi} \cdot  \vec{\pi}  \right]  \\ \nonumber
&& \times \frac{1}{6 v^2}  \left[ \left( \vec{\pi} \cdot \partial_{\mu} \vec{\pi} \right)^2  -  
\left(\vec{\pi} \cdot \vec{\pi} \right) \left(\partial^\mu\vec{\pi} \cdot \partial_{\mu} \vec{\pi} \right) \right] 
+ {\cal O} (\vec{\pi}^8)
\end{eqnarray}
  
The number of diagrams increases considerably
with the number of final state particles, making it impractical to perform an analytical computation. 
Hence we have implemented the Lagrangian given in Eq.(\ref{SILH}) in FormCalc \cite{formcalc} and MadGraph \cite{mg5} 
using FeynRules \cite{feynrules} (with UFO output \cite{ufo} for the higher dimensional operators).
We have also implemented this model in  CalcHEP \cite{calchep} 
using LanHEP package\cite{Semenov:2008jy} with the help of auxiliary fields.
This Lagrangian is equivalent to the usual linear sigma model by a field redefinition.

In the familiar case of $2 \rightarrow 2$ amplitudes the only kinematical dependence is on the Mandelstam variables $s$ and $t$.
For instance, denoting the goldstone bosons by their electric charge, the $00 \rightarrow +-$ amplitude, arising from only 4 diagrams, is given by:
\begin{equation}
{\cal M}_{00;+-} = \frac{s \left[ (1-a^2) s - m_h^2 \right]}{v^2 (s-m_h^2)} \underset{\tiny{ s \gg m_h^2}}{\longrightarrow} (1-a^2) \frac{s}{v^2} 
\end{equation}
Hence one can easily see that there is a violation of unitarity even with the presence of the Higgs boson if its coupling is not
SM-like, {\it i.e.}, $a \neq 1$. However, in the SM one obtains a constant amplitude at high energies, as expected.
 
The $2 \rightarrow 4$ amplitudes are much more complicated, containing of the order of 100 diagrams and depending on several 
combinations of the scalar products of the different 4-momenta involved. 
However, some of their properties can be demonstrated with the following simple example for a given point in phase space, where 
all the particles are in the same plane  (we will keep the ``pions" massless at the amplitude level 
since their masses are not relevant for issues of unitarity), in which case
we obtain
\begin{eqnarray}
&{\cal M}_{00;00+-}  \propto  \frac{1}{v^4} \left[ 72 s \left(13 a^4 - a^2 (7 b+5) -1 \right) + \right. \nonumber \\
 &\left. 3m_h^2 \left(1580 a^4 - 378 a^3 d_3 -3 a^2 (245 b + 131)-74\right) + \right. \nonumber \\
 &\left. \frac{m_h^4}{s} \left(9774 a^4 - 3087 a^3 d_3 - a^2 (4494 b + 1289) + 52 \right) + \right. \nonumber \\
 &\left.  \cdots \right]
\end{eqnarray}
It grows with $s$, as expected. However, in the SM ($a=b=d_3=1$) one obtains in the limit $s \gg m_h^2$:
\begin{equation}
{\cal M}_{00;00+-}  \propto \frac{1}{s} \frac{m_h^4}{ v^4} 
\end{equation}
and we explicitly see the strong cancellation where the first two terms in the amplitude vanish and the behaviour change  
from $s/v^4$ to $m_h^4/(s v^4)$, as anticipated.
The triple Higgs anomalous coupling parameterised by $d_3$ does not enter in the dominant contribution. 
In the following we will take $d_3 = 1$. There is no contribution from the couplings $d_4$ and $b_3$ for the above processes.
This result for the amplitude depends on the phase space configuration. The polynomials will be different 
but the features described above also happens at other phase space configurations and other channels.

For $2 \rightarrow 3$ processes a similar analysis can be performed.
For instance, again for a given configuration in phase space we find
\begin{eqnarray}
&{\cal M}_{00;hhh}  \propto  \frac{1}{4 v^3} \left[ s \left(-4 a^3 + 4 a b - 3 b_3)  \right) - \right. \nonumber \\
 &\left. m_h^2 \left( -8 a^3 + 8 a b + 3 b_3 \right) + \right. \nonumber \\
 &\left. \frac{4 m_h^4}{s} \left(a^3 + a b - 6 b_3 - 3 a^2 d_3 \right) + \cdots \right]
\end{eqnarray}
and
\begin{eqnarray}
&{\cal M}_{00;+-h}  \propto  \frac{a}{192 v^3} \left[ s \left(-1 + 2 a^2 - b \right) + \right. \nonumber \\
 &\left. \frac{m_h^2}{4} \left(-164 + 386 a^2 - 213 b - 9 a d_3 \right) - \right. \nonumber \\
 &\left. \frac{3 m_h^4}{2 s} \left(-262 + 291 a^2 - 93 b + 81 a d_3 \right) + \cdots \right]
\end{eqnarray}

We again find that for the SM the first two terms in these amplitudes vanish, as it should. Notice also that the ${\cal M}_{00;hhh}$ amplitude is sensitive to $b_3$, being the
lowest multiplicity process in which this happens.  In addition, as can be anticipated from the parity of the MCHM4 class of theories,
under which $\pi \rightarrow -\pi$ and $h \rightarrow -h$ \cite{CMPR},
the polynomial with largest growth in the $2 \rightarrow 3$ processes also vanish for the values of $a$, $b$ and $b_3$ that obey the MCHM4 
relations as in Eq.(\ref{coeffMCHM4}).

In summary, for  $a \neq 1$ and $b \neq 1$ the squared amplitude grows as $s^2$ instead of decreasing as $1/s^2$.
Therefore there is a large sensitivity of $2 \rightarrow 3$ and $2 \rightarrow 4$ processes to non-SM Higgs couplings compared to $2 \rightarrow 2$ processes, 
whereas the SM amplitude goes to a constant for large $s$. 
For the $2 \rightarrow 3$ processes, there is a also a suppression if the values predicted by the MCHM4 are used due to the
symmetry of the coset.  
In order to quantify this sensitivity we study directly the cross section for these processes in the next section.

\section{Sensitivity of $2\rightarrow 3, 4$ cross section to anomalous couplings}
In this section we analyse the cross section for the $2\rightarrow 3, 4$ processes  at the parton level for couplings
using the  Lagrangian Eq.(\ref{SILH}) implemented in CalcHEP. We studied several different channels but will
report only on the most representative ones. 

\begin{figure}[htb]
\begin{center}
\includegraphics[width=0.8\textwidth]{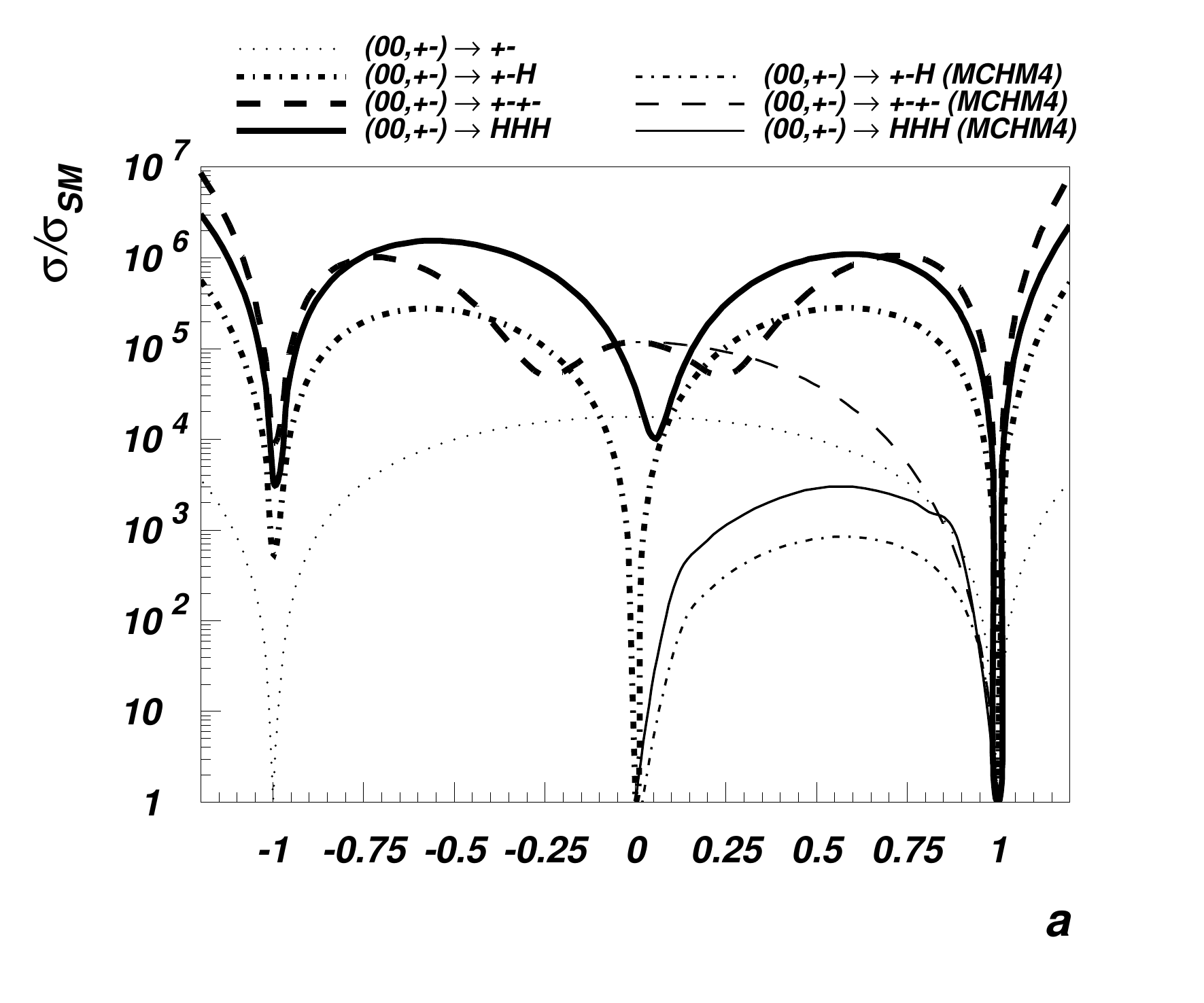}
\caption{Ratio of the SMEFF (thin lines) and MCHM4 (thick lines)  cross sections
to the SM one versus  $a$ parameter  at a fixed energy of $\sqrt{s} = 2$ TeV. 
The different channels are: $(00,+-) \rightarrow +-+-$ (dashed line),  
$(00,+-) \rightarrow +-h$ (dot-dashed line),
$(00,+-) \rightarrow hhhh$ (solid line),
and  $(00,+-) \rightarrow +-$ (dotted line) for comparison.
The notation $(00,+-)$ indicates that both $00$ and $+-$ initial states were taken into account.\label{CrossRatio}
}
\end{center}
\end{figure}

In order to show examples of the enhancements that result from the anomalous
Higgs couplings, we compute cross sections with a Higgs mass $m_h = 125$ GeV. We implement a cut in the invariant mass of the
final state pions $m_{+-} > 200$ GeV, such that the Higgs is never on-mass-shell when coupled to two pions, as is the case
in the SM (the Higgs does not decay to a on-shell pair of gauge bosons).
We show in Fig.\ref{CrossRatio} the ratio of the cross section as a function of $a$ (keeping the other parameters fixed in one case and, in the case of MCHM4, 
changing them according to Eq.(\ref{coeffMCHM4}))  
to the SM cross section 
at a fixed center-of-mass energy of $\sqrt{s} = 2$ TeV. 
Several channels are shown, such as $(00,+-) \rightarrow +-, +-h, hhh, +-+-$.
The notation $(00,+-)$ indicates that both $00$ and $+-$ initial states have been taken into account
in the cross section.
Thick lines show results for different values of $a$ but keeping the other parameters fixed at their SM values
whereas thin lines show results when the other parameters are changed according to the MCHM4 where, in addition
to the relations given in Eq.(\ref{coeffMCHM4}), one assumes $d_3 = \sqrt{ 1 - \xi}$.
Of course not all the values 
of the anomalous couplings are allowed: this plot is only meant for illustrative purposes. 

Large enhancements of the order of $10^6$ with respect to the SM value are easily obtained even for small deviations (as small as $10 \%$) 
of the couplings from their SM values.
The behaviour of the curves are easily understandable: the 4, 3 and 2 dips in the cross section versus $a$ for $2 \rightarrow 4$, $2\rightarrow 3$ and $2 \rightarrow 2$ 
are due to the
$4^{th}$, $3^{rd}$ and $2^{nd}$ order polynomials in $a$ in the amplitudes. One can see that the enhancements in $2 \rightarrow 2$ processes are modest compared 
to processes with higher multiplicities, at least at 2 TeV. For the MCHM4 case there is a suppression in the $2 \rightarrow 3$ process, as expected from the parity
symmetry of the coset. Since the MCHM4 always predict smaller deviations,
in what follows we will consider the more optimistic case where the parameter $a$ can be the only one different from the SM values.

\begin{figure}[htb]
\begin{center}
\vskip -0.8cm
\includegraphics[width=0.7\textwidth]{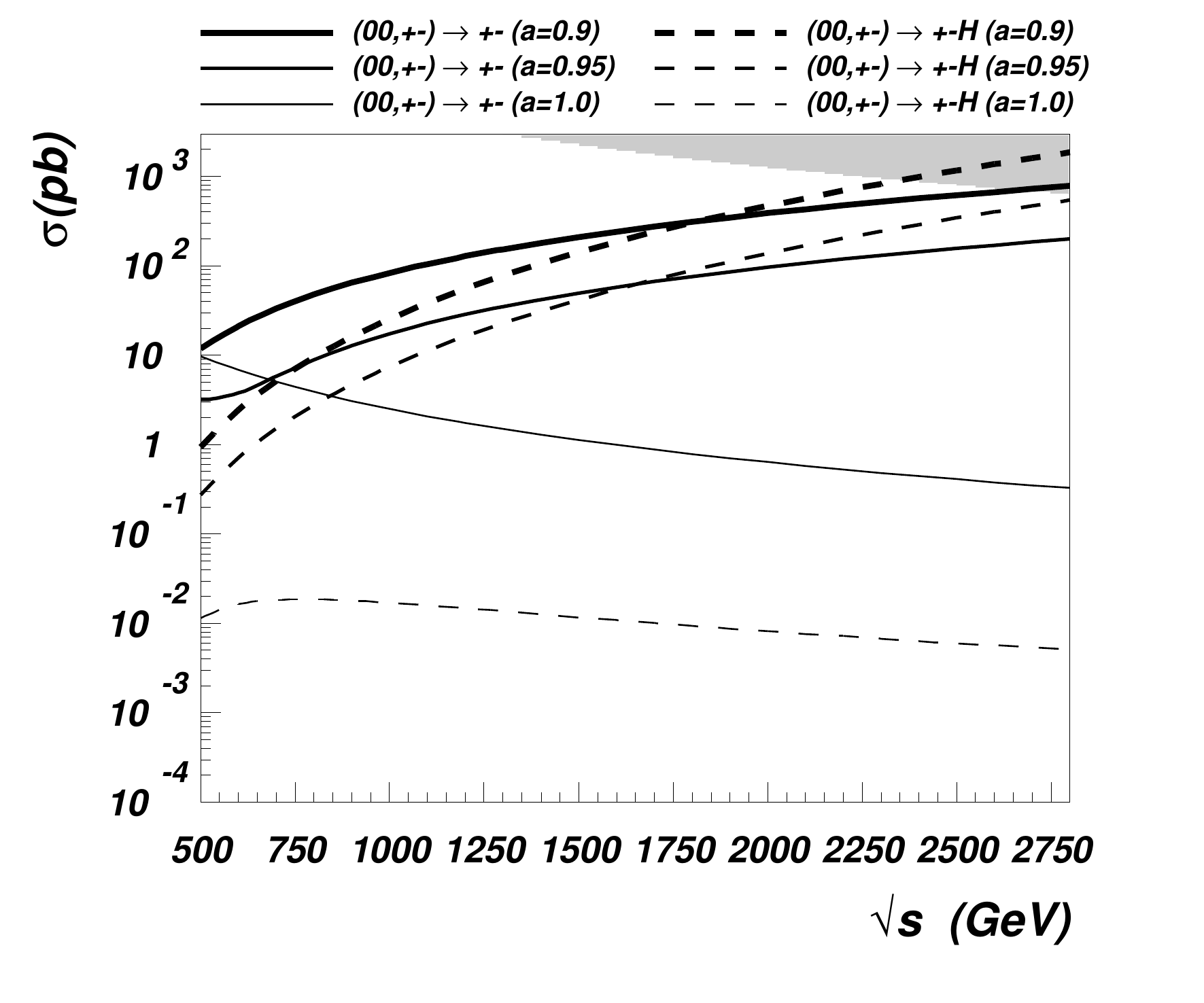}
\vskip -0.8cm
\includegraphics[width=0.7\textwidth]{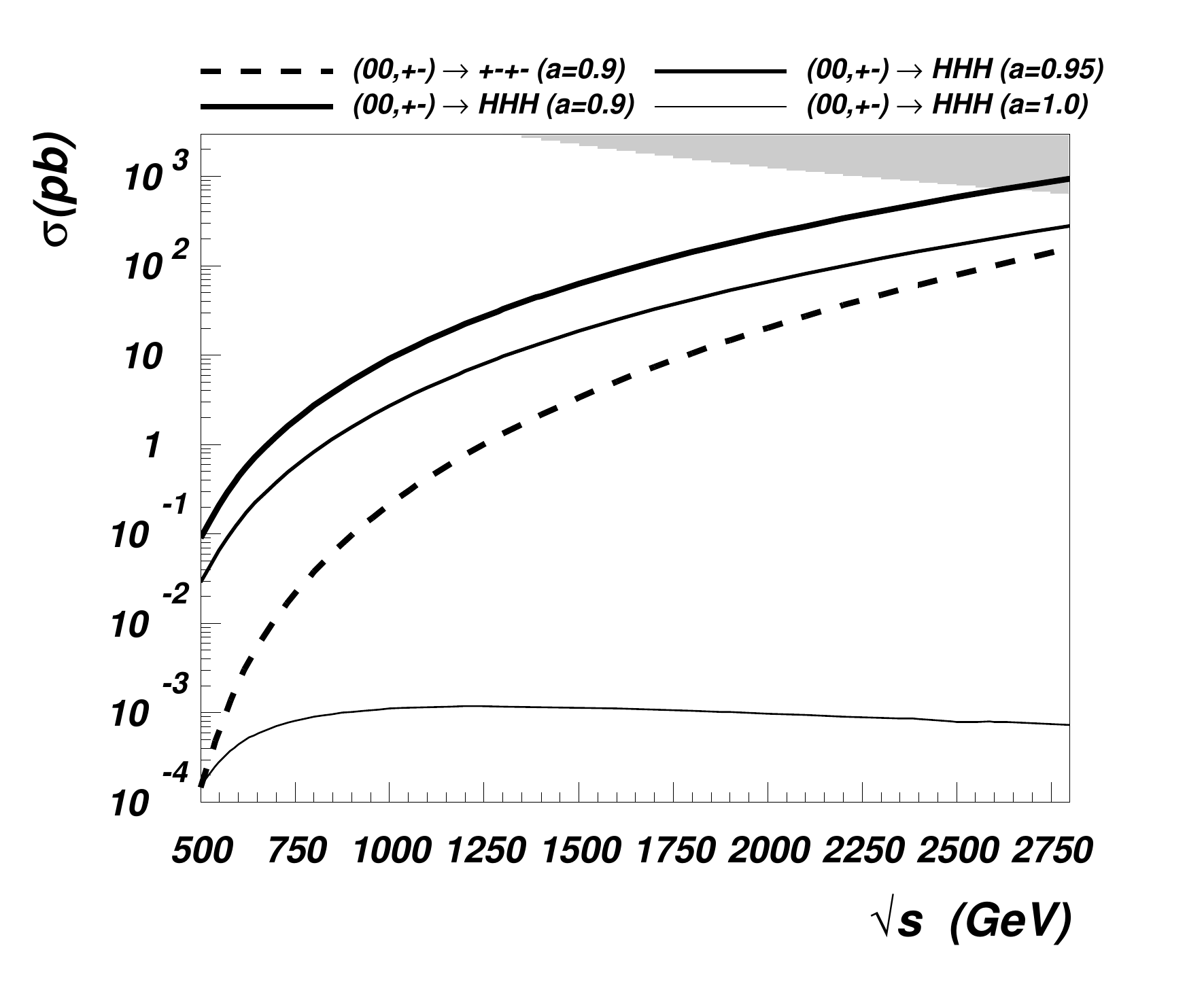}
\vskip -0.8cm
\caption{Comparison among cross sections as a function of the center-of-mass energy for processes with $2, 3$ and $4$ particles in the final state.
In the top plot,
the solid lines are for $(00,+-) \rightarrow +-$ for $a=0.9$ (thick), $a=0.95$ (medium thick) and $a=1$ (thin). Dashed lines are for $(00,+-) \rightarrow +-h$, with
the same pattern for the thickness of the lines. 
In the bottom  plot, the same pattern of lines show the results $(00,+-) \rightarrow hhh$ and the process $00 \rightarrow +-+-$ is shown as a dashed line for $a=0.9$.  
In these plots only $a$ deviates from the SM value.
The unitarity bound is shown as a shaded area in the top right corner.\label{test}
}
\end{center}
\end{figure}

Next we study the growth with center-of-mass energy of the cross section for different multiplicities for 
a few values of the anomalous coupling, namely $a=0.9$, $0.95$ and $1$ (SM), 
keeping the other couplings at their SM values. 
We believe that these values of the anomalous
couplings can be representative of the behaviour of the cross sections, that is, we expect the same
order-of-magnitude enhancements if the other couplings are also anomalous (but without obeying the MCHM4 relations).
In Fig.~\ref{test} we present a comparison of the cross section as a function of energy 
among representative processes with $2, 3$ and $4$ particles in the final state, for different values of the anomalous coupling.
We also  show the unitarity limit Eq.(\ref{unitarity}).
A few comments are in order. The SM cross section quickly stabilises at a small value, which depends on the specific
process (of the order of $10^{-3}$ pb,  $10^{-2}$ pb and  $10^{-1}$ pb for $00 \rightarrow  hhh$,  $(00,+-) \rightarrow + - h$ and $(00,+-) \rightarrow +-$, respectively).
It is not surprising that in the non-SM case the cross sections grow very fast with energy, reaching up to order of $100$ pb
and violating unitarity at center-of mass energies of the order of a few TeV. 
It is also anticipated that larger multiplicity processes, in the absence of a complete cancellation mechanism, grow faster due to phase space.
However, what is somewhat unexpected is the energy scale at which multiparticle cross sections become comparable to $2 \rightarrow 2$ processes.
In the examples shown the $2 \rightarrow 3$ start to overcome $2 \rightarrow 2$ at energies of ${\cal O} (1 \mbox{TeV})$.
This may be signalling the onset of nonperturbative behaviour well
before the unitarity bound is reached. It is not clear whether new physics must come in at these scales, as for instance the appearance of new resonances.
In this work we assume that this is not the case.
We also checked that the $2 \rightarrow 4$ process grows very rapidly for $a \neq 1$, 
but since it starts out very suppressed it surpasses the
$2 \rightarrow 2$ only at very high energies, of the order of ${\cal O} (5 \mbox{TeV})$.

\section{Cross sections in the SM with anomalous Higgs couplings}
So far we have only analysed the scattering of the longitudinally polarised gauge bosons.
Since it is difficult to separate out the contributions from these polarisations in an experimental setting,
it is important to understand how the large enhancements found will affect 
the corresponding unpolarised cross section.
In order to do so, we use the Lagrangian Eq.(\ref{SILH}) promoting the partial derivatives to 
full covariant derivatives and adopt the unitary gauge ($U = \mathbf{1}$).
As an illustration, we fix the partonic center-of-mass energy at 2 TeV and compare the cross sections
for the longitudinally polarised gauge bosons with the full SM in Table~\ref{cs1}. We still keep the notation $0,+,-$
to indicate longitudinally polarised gauge bosons and $Z, W^{\pm}$ to denote the unpolarised gauge bosons.

\begin{table}[h]
\begin{center}
\begin{tabular}{|l|l|l|}
\hline
channel & $a=b=1$ (SM) & $a=0.9; b=1$  \\
\hline \hline
$00\rightarrow +- $ & 0.13 & 295  \\
\hline
$ZZ\rightarrow W^+ W^-$ & 610   & 655    \\
\hline 
$00\rightarrow +- h$ & $2.0 \times 10^{-3}$  & 350  \\
\hline
$ZZ\rightarrow  W^+ W^- h$ & 10.9 & 46.2   \\
\hline \hline
$00\rightarrow hh$ & 0.18 & 158  \\
\hline
$ZZ\rightarrow hh$ & 7.61   & 15.7    \\
\hline 
$00\rightarrow hhh$ & $4.9 \times 10^{-4}$  & 112  \\
\hline
$ZZ\rightarrow  hhh$ & $4.65 \times 10^{-2}$& 13.6   \\
\hline
\end{tabular}
\caption{Comparison of $2 \rightarrow 2$ and $2 \rightarrow 3$ cross sections (in picobarns) at $\sqrt{s} = 2$ TeV.\label{cs1}}
\end{center}
\end{table}

Notice that the processes with longitudinal polarisations
are subdominant in the SM. However, as we discussed above, they are greatly enhanced with
small deviations of the couplings and actually dominate the cross sections. 
The results are consistent with the fact that $\sigma_{\mbox{{\small all}}} \approx  \sigma_{LL}/9$
when the contribution from longitudinal polarisations is dominant.
The enhancements are larger when the final state multiplicity 
is larger, as expected. For instance, $\sigma(00 \rightarrow +-h) > \sigma(00 \rightarrow +-)$
for $a=0.9$ and $b=1$. When all polarisations are included, the contribution from transverse polarisations
can mask the increase in the cross section for the longitudinally polarised gauge bosons. This can be seen in the case of $ZZ \rightarrow W^+ W^-$,
where the total increase in the cross section is less than $10 \%$. On the other hand, in cases where the contributions
from the transverse polarisations are not large, as in the case of $ZZ \rightarrow hhh$, enhancements of ${ \cal O} (10^3)$
can be obtained. Therefore, multiple Higgs production offers the best channels to study anomalous couplings.
For the unpolarised case, the cross section for $ZZ \rightarrow W^+ W^-$ is still at least one order of magnitude larger than
the typical $2 \rightarrow 3$ processes but $ZZ \rightarrow hh$ is of the same order as the $ZZ \rightarrow hhh$ cross section.
In the next Section we discuss the impact of these results for the LHC and future colliders.

\section{Impact of multiparticle production at the LHC and future colliders}

In order to estimate the impact of these enhancements found at the parton level arising from anomalous Higgs couplings, 
we have performed a full calculation of $pp \rightarrow jj + X$, where $j = u, \bar{u}, d, \bar{d}, s, \bar{s}$ and 
$X= W^+ W^-,  W^+ W^- h, hhh$ at the LHC (for $\sqrt{s} = 14$ and $33$ TeV) using Madgraph5 (v1.4.8). 
We use CTEQ6L1 parton density function for the evaluation of the 
tree-level cross sections with the QCD scale equal to $M_Z$.
The selection and acceptance cuts include
the  requirement of  two  jets with  $P_{T_j} > 30$ GeV with 
$|\eta_j|<5$ separated with $\Delta R=\sqrt{\Delta \phi_{jj}^2+\Delta \eta_{jj}^2}>0.4$.
Besides the cross sections evaluated for the cuts above
we have evaluated another set of the cross sections for 
an additional  cut to select the vector boson fusion process by requiring
each jet to be quite energetic with $E_j>300$ GeV as well as a  large rapidity gap 
between the two jets  $|\Delta \eta_{jj}| > 4$
(see e.g. \cite{He:2007ge} for detailed motivation of  this choice).
In Table~\ref{cs2} we present  the results with and without the vector boson fusion cut.
\begin{table}[htb]
\begin{center}
\begin{tabular}{|l|l|l|l|l|}
\hline
        & \multicolumn{2}{|c|}{14 TeV} & \multicolumn{2}{|c|}{33 TeV}\\
\cline{2-5}
Process & \multicolumn{2}{|c|}{with (without) VBF cuts} & \multicolumn{2}{|c|}{with (without) VBF cuts} \\
\cline{2-5}
	& a=1.0& a=0.9 & a=1.0 & a=0.9\\ 
	& b=1.0& b=1.0 & b=1.0 & b=1.0\\ 
\hline
        & & & &\\
$pp\to jj W^+W^-$  &$95.2$            & $99.3  $                & $512$            &$540$\\	
                  &$(1820)$             & $(1700) $                & $(5120)$            &$(5790)$\\	

        & & & &\\
\hline
        & & & &\\
$pp\to jj W^+W^-h$ &$0.011  $ & $0.0088    $& $0.0765 $&$0.0626$\\	
                  &$(0.206)$ & $(0.172) $& $(0.914)$&$(0.758)$\\	

        & & & &\\
\hline
        & & & &\\
$pp\to jj hhh$ &$1.16\times 10^{-4}  $ & $0.0566   $& $0.00115$&$1.85$\\	
              &$(3.01\times 10^{-4})$ & $(0.0613) $& $(0.00165)$&$(1.46)$\\	
        & & & & \\
\hline
\end{tabular}
\caption{Cross section (in fb) for $pp\to jj W^+W^-$, $pp\to jj W^+W^-h$ and $pp\to jj hhh$ processes evaluated with Madgraph5.\label{cs2}}
\end{center}
\end{table}

\begin{figure}[htb]
\begin{center}
\includegraphics[width=0.7\textwidth]{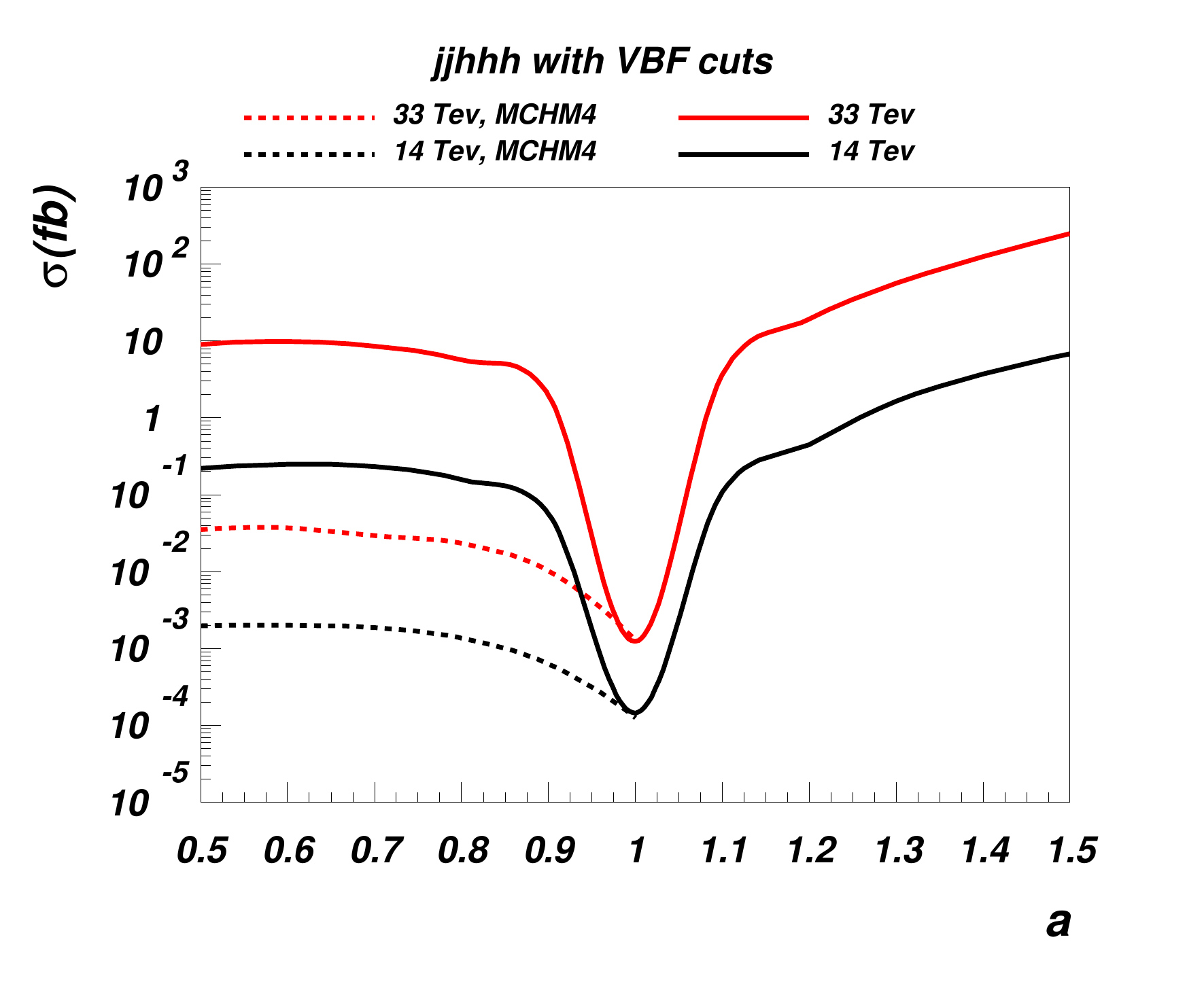}
\caption{Cross section for triple Higgs production $pp \rightarrow j j hhh$ with VBF cuts as a function of the anomalous coupling $a$ for LHC14 (dark lines) and LHC33 (light lines). 
Dashed lines are for other parameters fixed to SM values and solid lines are for parameters given by MCHM4 relations.\label{jjhhh}
}
\end{center}
\end{figure}

One can notice that when there are gauge bosons in the final state the cross section actually decreases for most cases
with $a=1$ versus $a=0.9$ ones. This is because
we chose in our example $a < a_{SM} = 1$ and since
the transverse polarisations dominate the cross section, reducing the coupling $a$ results in a smaller cross section. However, in the case
of triple Higgs production, the enhancements are substantial: 
roughly a factor of 500 for $\sqrt{s} = 14$ TeV (LHC14) and $1600$ for $\sqrt{s} = 33$ TeV (LHC33), with VBF cuts.
We show in Fig.~\ref{jjhhh} the results for the triple Higgs production cross section for both LHC14 and LHC33 for the large range of  anomalous  coupling $a$.
The enhancements with respect to the SM case $a=1$ are large and don't change significantly once $|\Delta a /a| > 0.1$.

Though the enhancement can be as large as $10^5$
for $a=1.5$, the absolute value of
the respective cross sections  are quite low (about 10 fb for $\sqrt{s} = 14$~TeV with VBF cuts)
making  the study of these processes challenging  at the LHC.
A dedicated analysis (which are outside the scope of this paper)
are necessary to understand the LHC or LHC33 sensitivity to the processes above.
However we can already estimate that if such sensitivity is possible it can take place only
at high luminocities and/or high energies for quite large values of the $a$ parameter.
A detailed analysis of these processes at future 
$e^+ e^-$ colliders is being performed in \cite{grojeanetal}.

\section{Conclusion}
In this Letter we have studied multiparticle production in models with anomalous Higgs couplings, such as the composite Higgs models.
The modified couplings result in a partial unitarization of the scattering amplitudes. 
We found that, due to the stronger cancellations in the corresponding amplitudes compared to the usual $2 \rightarrow 2$
processes, very large enhancements with respect to the SM, as large as ${\cal O} (10^6)$, can arise at the parton level in the cross section of
longitudinally polarised gauge bosons, even for small deviations of the 
couplings from their SM values. The cross sections grow faster with energy for larger multiplicities, as expected from naive
phase space considerations. We pointed out that some $2 \rightarrow 3$ processes become as important as $2 \rightarrow 2$ processes 
for relatively low energies, of the order of a TeV, signalling the onset of nonperturbative effects.
When accounting for the contributions from the transverse polarisations, the enhancements are somewhat diluted but remain
important in some processes, especially triple Higgs production.

However, we showed with a realistic calculation that even with these large enhancements the search for multiparticle processes
will remain a challenge for the foreseeable future. 
On the other hand,  multiple gauge and Higgs boson 
production receives  large enhancements in the case of anomalous Higgs boson couplings, and its study could be an important part of  future experimental programs aiming at  understanding 
underlying theory beyond the Standard Model. 

\section*{Acknowledgements}
It is a pleasure to thank C. Grojean, H.-J. He and C. Quigg for helpful discussions.
RR and ACAO thank the Theory Division at CERN for the hospitality during the
development of this work.
RR was supported by a FAPESP grant 2011/10199-3, ACAO by a CAPES fellowship, and MCT by a STFC studentship grant.
RR thanks IFT-UNESP for granting a sabbatical leave.
AB acknowledges partial support from the STFC Consolidated ST/J000396/1
grant as well as from  Royal Society grants JP090598 and JP090146. 


\end{document}